\documentstyle[aaspp4,epsfig]{article}
\slugcomment{OU-TAP 97: ApJL, in press}
\begin{document}

\title{Cosmological Implications of the Fundamental Relations of X-ray
  Clusters} \author{Yutaka Fujita and
  Fumio Takahara} \affil{Department of Earth and Space Science,
  Graduate School of Science, Osaka University, Machikaneyama-cho,
  Toyonaka, Osaka, 560-0043, Japan}
\authoremail{fujita@vega.ess.sci.osaka-u.ac.jp}

\begin{abstract}

  Based on the two-parameter family nature of X-ray clusters of galaxies
  obtained in a separate paper, we discuss the formation history of
  clusters and cosmological parameters of the universe. Utilizing the
  spherical collapse model of cluster formation, and assuming that the
  cluster X-ray core radius is proportional to the virial radius at the
  time of the cluster collapse, the observed relations among the
  density, radius, and temperature of clusters imply that cluster
  formation occurs in a wide range of redshift. The observed relations
  favor the low-density universe. Moreover, we find that the model of
  $n\sim -1$ is preferable.

\end{abstract}


\section{Introduction}

Galaxy clusters are the largest virialized objects in the universe and
provide useful cosmological probes, since several properties of clusters
are strongly dependent on the cosmological parameters. For example, the
statistics of X-ray clusters can serve as an excellent probe of
cosmology. The abundance of clusters and its redshift evolution can be
used to determine the cosmological density parameter, $\Omega_0$, and
the rms amplitude of density fluctuations on the fiducial scale $8
h^{-1}$ Mpc, $\sigma_8$ (e.g. White, Efstathiou, \& Frenk
\markcite{wef1993}1993 ; Eke, Cole, \& Frenk \markcite{ecf1996}1996 ;
Viana \& Liddle \markcite{vl1996}1996 ; Bahcall, Fan, \& Cen
\markcite{bfc1997}1997 ; Fan, Bahcall, \& Cen
\markcite{fbc1997}1997). Moreover, temperature and luminosity function
of X-ray clusters is also used for a cosmological probe. Taking account
of the difference between cluster formation redshift and observed
redshift, Kitayama \& Suto \markcite{ks1996}(1996) computed a
temperature and luminosity function semi-analytically; comparing the
predicted temperature ($T$) and luminosity ($L_{\rm X}$) function with
the observed ones, they conclude that $\Omega_0 \sim 0.2- 0.5$ and $h
\sim 0.7$. However, one tenacious problem in such investigations is the
discrepancy of $L_{\rm X}-T$ relation between observations and simple
theoretical prediction. This relation should also be an important probe
of the formation history and cosmology.

In a separate paper (Fujita \& Takahara \markcite{ft1999}1999;
hereafter Paper I), we have shown that the clusters of galaxies populate
a planar distribution in the global parameter space $(\log \rho_0, \log
R, \log T)$, where $\rho_0$ is the central gas density, and $R$ is the
core radius of clusters of galaxies. This 'fundamental plane' can be
described as
\begin{equation}
  \label{eq:fplane}
  X=\rho_0^{0.47} R^{0.65} T^{-0.60}=constant \:.
\end{equation}
We thus find that clusters of galaxies form a two-parameter family. The
minor and major axes of the distribution are respectively given by
\begin{equation}
  \label{eq:Y}
  Y=\rho_0^{0.39} R^{0.46} T^{0.80} \:,
\end{equation}
\begin{equation}
  \label{eq:Z} 
  Z=\rho_0^{0.79} R^{-0.61}T^{-0.039} \:.
\end{equation}
The scatters of observational data in the directions of $Y$ and $Z$ are 
$\Delta \log Y = 0.2$ and $\Delta \log Z = 0.5$, respectively (Paper I).
The major axis of this 'fundamental band' ($Z$) is nearly parallel to
the $\log R-\log \rho_0$ plane, and the minor axis ($Y$) describes the
$L_{\rm X}-T$ relation.

In this Letter, we discuss cosmological implications of the relations we
found in Paper I, paying a particular attention to the two-parameter
family nature of X-ray clusters and to the difference between cluster
formation redshift and observed redshift. In \S\ref{sec:evo}, we use
spherical collapse model to predict the formation history of clusters of
galaxies, and in \S\ref{sec:dis}, we predict the observable distribution
of X-ray clusters.

\section{Formation History of Clusters of Galaxies}
\label{sec:evo}

In order to explain the observed relations between the density, radius,
and temperature of clusters of galaxies, we predict them for a flat and
an open universe theoretically with the spherical collapse model (Tomita
\markcite{t1969}1969; Gunn \& Gott \markcite{gg1972}1972). For
simplicity, we do not treat vacuum dominated model in this paper. For a
given initial density contrast, the spherical collapse model predicts
the time at which a uniform spherical overdense region, which contains
mass of $M_{\rm vir}$, gravitationally collapses. Thus, if we specify
cosmological parameters, we can obtain the collapse or formation
redshifts of clusters. Moreover, the model predicts the average density
of the collapsed region $\rho_{\rm vir}$.

In Paper I, we showed that the observed fundamental band are described
by the two independent parameters $M_{\rm vir}$ and $\rho_{\rm vir}$. In
particular, the variation of $\rho_{\rm vir}$ is basically identified
with the scatter of $Z$. Since the spherical collapse model treats
$\rho_{\rm vir}$ and $M_{\rm vir}$ as two independent variables, it can
be directly compared with the observed fundamental band, as long as we
assume that the core radius and the mass of core region are respectively
a fixed fraction of the virial radius and that of the virial mass at the
collapse redshift, as is adopted in this paper. Although the model may
be too simple to discuss cosmological parameters quantitatively, it can
plainly distinguish the relations between the density, radius, and
temperature in a low-density universe from those in a flat universe, as
shown below.

For the spherical model, the virial density of a cluster is $\Delta_c$
times the critical density of a universe at the redshift when the
cluster collapsed ($z_{\rm coll}$). It is given by
\begin{equation}
  \label{eq:density}
  \rho_{\rm vir} = \Delta_c \rho_{\rm crit}(z_{\rm coll})
           = \Delta_c \rho_{\rm crit,0}E(z_{\rm coll})^2
           = \Delta_c \frac
             {\Omega_0 \rho_{\rm crit,0} (1+z_{\rm coll})^3}
             {\Omega(z_{\rm coll})} \:,
\end{equation}
where $\Omega(z)$ is the cosmological density parameter, and $E(z)^2 =
\Omega_0 (1+z)^3/\Omega(z)$. The index 0 refers to the values at
$z=0$. Note that the redshift-dependent Hubble constant can be written
as $H(z) = 100 h E(z) \rm\; km\; s^{-1}\; Mpc^{-1}$. We fix $h$ at 0.5.
In practice, we use the fitting formula of Bryan \& Norman
\markcite{bn1998}(1998) for the virial density:
\begin{equation}
  \label{eq:delta}
  \Delta_c = 18\pi^2 + 60 x - 32 x^2  \;,
\end{equation}
where $x = \Omega(z_{\rm coll})-1$. 

It is convenient to relate the collapse time in the spherical model with
the density contrast calculated by the linear theory. We define the
critical density contrast $\delta_c$ that is the value, extrapolated to
the present time ($t=t_0$) using linear theory, of the overdensity which
collapses at $t=t_{\rm coll}$ in the exact spherical model. It is given
by
\begin{eqnarray}
  \label{eq:crit}
  \delta_c(t_{\rm coll}) 
       &=& \frac{3}{2}D(t_0)\left[
           1+\left(\frac{t_\Omega}{t_{\rm coll}}\right)^{2/3}\right]
               \;\;\;\; (\Omega_0 < 1) \\
       &=& \frac{3(12\pi)^{2/3}}{20}
           \left(\frac{t_0}{t_{\rm coll}}\right)^{2/3}
               \;\;\;\; (\Omega_0 = 1) 
\end{eqnarray}
(Lacey \& Cole \markcite{lc1993}1993), where $D(t)$ is the linear
growth factor given by equation (A13) of Lacey \& Cole
\markcite{lc1993}(1993) and $t_\Omega = \pi
H_0^{-1}\Omega_0(1-\Omega_0)^{-3/2}$. 

For a power-law initial fluctuation spectrum $P\propto k^n$, the rms
amplitude of the linear mass fluctuations in a sphere containing an
average mass $M$ at a given time is $\delta \propto M^{-(n+3)/6}$. Thus,
the virial mass of clusters which collapse at $t_{\rm coll}$ is related
to that at $t_0$ as
\begin{equation}
  \label{eq:mass}
  M_{\rm vir}(t_{\rm coll}) 
     =M_{\rm vir}(t_0)\left[
  \frac{\delta_c(t_{\rm coll})}{\delta_c(t_0)}\right]^{-6/(n+3)}
                  \;.
\end{equation}
Here, $M_{\rm vir}(t_0)$ is regarded as a variable because actual
amplitude of initial fluctuations has a distribution.  We relate
$t=t_{\rm coll}$ to the collapse or formation redshift $z_{\rm coll}$,
which depends on cosmological parameters. Thus, $M_{\rm vir}$ is a
function of $z_{\rm coll}$ as well as $M_{\rm vir}(t_0)$. This means
that for a given mass scale $M_{\rm vir}$, the amplitude takes a range
of value, and thus spheres containing a mass of $M_{\rm vir}$ collapse
at a range of redshift. In the following, the slope of the spectrum is
fixed at $n = - 1$, unless otherwise mentioned. It is typical in the
scenario of standard cold dark matter for a cluster mass range.

The virial radius and temperature of a cluster are then calculated by
\begin{equation}
  \label{eq:rad}
  r_{\rm vir} = \left(\frac{3M_{\rm vir}}
               {4\pi \rho_{\rm vir}}\right)^{1/3} \:,
\end{equation}
\begin{equation}
  \label{eq:temp}
  T_{\rm vir} = \frac{\mu m_{\rm H}}{3 k_{\rm B}}\frac{G M_{\rm
      vir}}{r_{\rm vir}} \:,
\end{equation}
where $\mu (=0.6)$ is the mean molecular weight, $m_{\rm H}$ is the
hydrogen mass, $k_{\rm B}$ is the Boltzmann constant, and $G$ is the
gravitational constant.

\section{Results and Discussion}
\label{sec:dis}

Since equations (\ref{eq:density}) and (\ref{eq:mass}) show that
$\rho_{\rm vir}$ and $M_{\rm vir}$ are the functions of $z_{\rm coll}$
for a given $M_{\rm vir}(t_0)$, the virial radius $r_{\rm vir}$ and
temperature $T_{\rm vir}$ are also the functions of $z_{\rm coll}$ for a
given $M_{\rm vir}(t_0)$ (equations [\ref{eq:rad}] and [\ref{eq:temp}]). 
Thus, by eliminating $z_{\rm coll}$, the relations among them can be
obtained.
Since observational values reflect mainly the structures of core region
while the theory predicts average values within the virialized region,
we must specify the relation between the observed values $(\rho_0, R,
T)$ and theoretically predicted values $(\rho_{\rm vir}, R_{\rm vir},
T_{\rm vir})$. Since we assume that mass distribution of clusters is
similar, $r_{\rm vir} \propto R$ and $T_{\rm vir}= T$, emphasizing that
$r_{\rm vir}$ is the virial radius when the cluster collapsed (see
Salvador-Sol\'{e}, Solanes, \& Manrique \markcite{ssm1998}1998). In this
case, the typical gas density of clusters has the relation $\rho_0
\propto f\rho_{\rm vir}$, where $f$ is the baryon fraction of the
cluster. Since it is difficult to predict $f$ theoretically, we assume
$f\propto M_{\rm vir}^{0.4}$, which is consistent with the observations
and corresponds to $X\sim constant$ (Paper I). For definiteness, we
choose
\begin{equation}
  \label{eq:gasden}
  f = 0.25\left(\frac{M_{\rm vir}}{10^{15}\rm\;M_{\sun}}\right)^{0.4} 
                  \:.
\end{equation}
Figure 1 shows the predicted relations between $f\rho_{\rm vir}$,
$r_{\rm vir}$, and $T_{\rm vir}$ for $\Omega_0 = 0.2$ and $1$. Since we
are interested only in the slope and extent of the relations, we do not
specify $M_{\rm vir}(t_0)$ exactly. Moreover, since $M_{\rm vir}(t_0)$
has a distribution, we calculate for $M_{\rm vir}(t_0)= 10^{16}\rm\;
M_{\sun}$ and $5\times 10^{14}\rm\; M_{\sun}$. The lines in Figure 1
correspond to the major axis of the fundamental band or $Z$ because they
are the one-parameter family of $\rho_{\rm vir}$.  The width of the
distribution of $M_{\rm vir}(t_0)$ represents the width of the band or
$Y$. The observational data of clusters are expected to lie along these
lines according to their formation period.  However, when $\Omega_0 =
1$, most of the observed clusters collapsed at $z\sim 0$ because
clusters continue growing even at $z=0$ (Peebles
\markcite{p1980}1980). Thus, the cluster data are expected to be
distributed along the part of the lines close to the point of $z_{\rm
coll}=0$ (segment ab). In fact, Monte Carlo simulation done by Lacey \&
Cole \markcite{lc1993}(1993) shows that if $\Omega_0 = 1$, most of
present clusters ($M_{\rm vir}\sim 10^{15}\rm\: M_{\sun}$) should have
formed in the range of $z_{\rm coll}<0.5$ (parallelogram abcd).  When
$\Omega_0 = 0.2$, the growing rate of clusters decreases and cluster
formation gradually ceases at $z \lesssim 1/\Omega_0-1$ (Peebles
\markcite{p1980}1980). Thus, cluster data are expected to be distributed
between the point of $z_{\rm coll}=0$ and $z_{\rm coll} = 1/\Omega_0-1$
and should have a two-dimensional distribution (parallelogram ABCD).

The observational data are also plotted in Figure 1. The data are the
same as those used in Paper I.  For definiteness, we choose $r_{\rm
vir}=8 R$ and $f\rho_{\rm vir}=0.06\rho_0$. The figure shows that the
model of $\Omega_0 = 0.2$ is generally consistent with the observations.
The slopes of lines both in the model of $\Omega_0 = 0.2$ and
$\Omega_0=1$ seem to be consistent with the data, although the model of
$\Omega_0 = 0.2$ is preferable. However, the model of $\Omega_0 = 1$ is
in conflict with the extent of the distribution because this model
predicts that the data of clusters should be located only around the
point of $z_{\rm coll}=0$ in Figure 1.

Finally, we comment on the case of $n=-2$, which is suggested by the
analysis based on the assumption that clusters have just formed when
they are observed (e.g. Markevitch \markcite{m1998}1998). Figure 2 is
the same as Figure 1b, but for $n= -2$. The theoretical predictions are
inconsistent with the observational results, because the theory predicts
rapid evolution of temperatures; there should be few clusters with
high-temperature and small core radius.

The results of this Letter suggest $\Omega_0 < 1$, and that the clusters
of galaxies existing at $z\sim 0$ include those formed at various
redshifts. We also show that the model of $n\sim -1$ is favorable.
Importantly, the location of a cluster in these figures tells us
its formation redshift. In order to derive the cosmological parameters
more quantitatively, we should consider merging history of clusters and
predict the mass function of clusters for each $z$.

\acknowledgments

This work was supported in part by the JSPS Research Fellowship for
Young Scientists.

\newpage

\newpage 

\begin{figure}
\centering \epsfig{figure=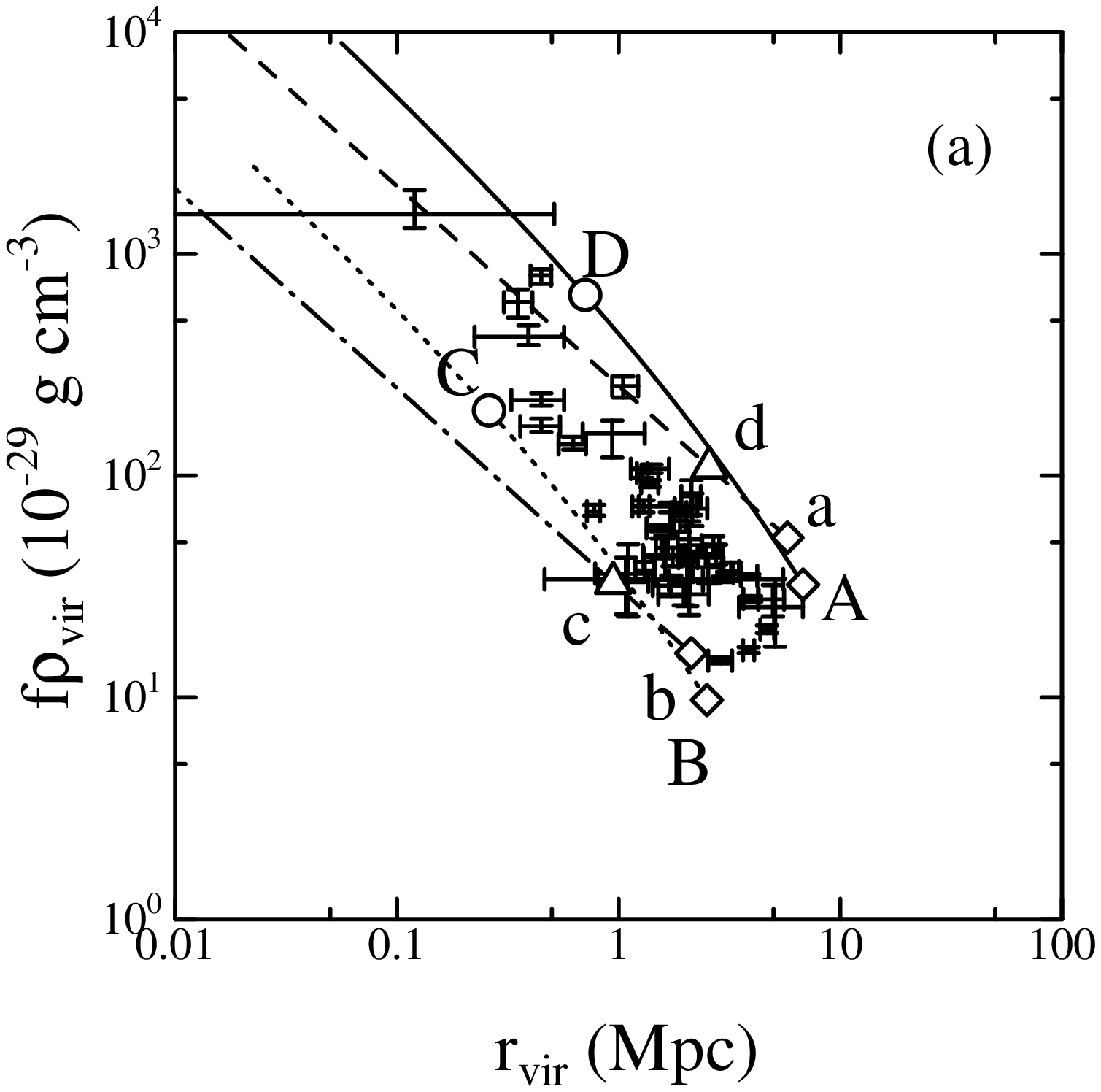, width=7cm}
\centering \epsfig{figure=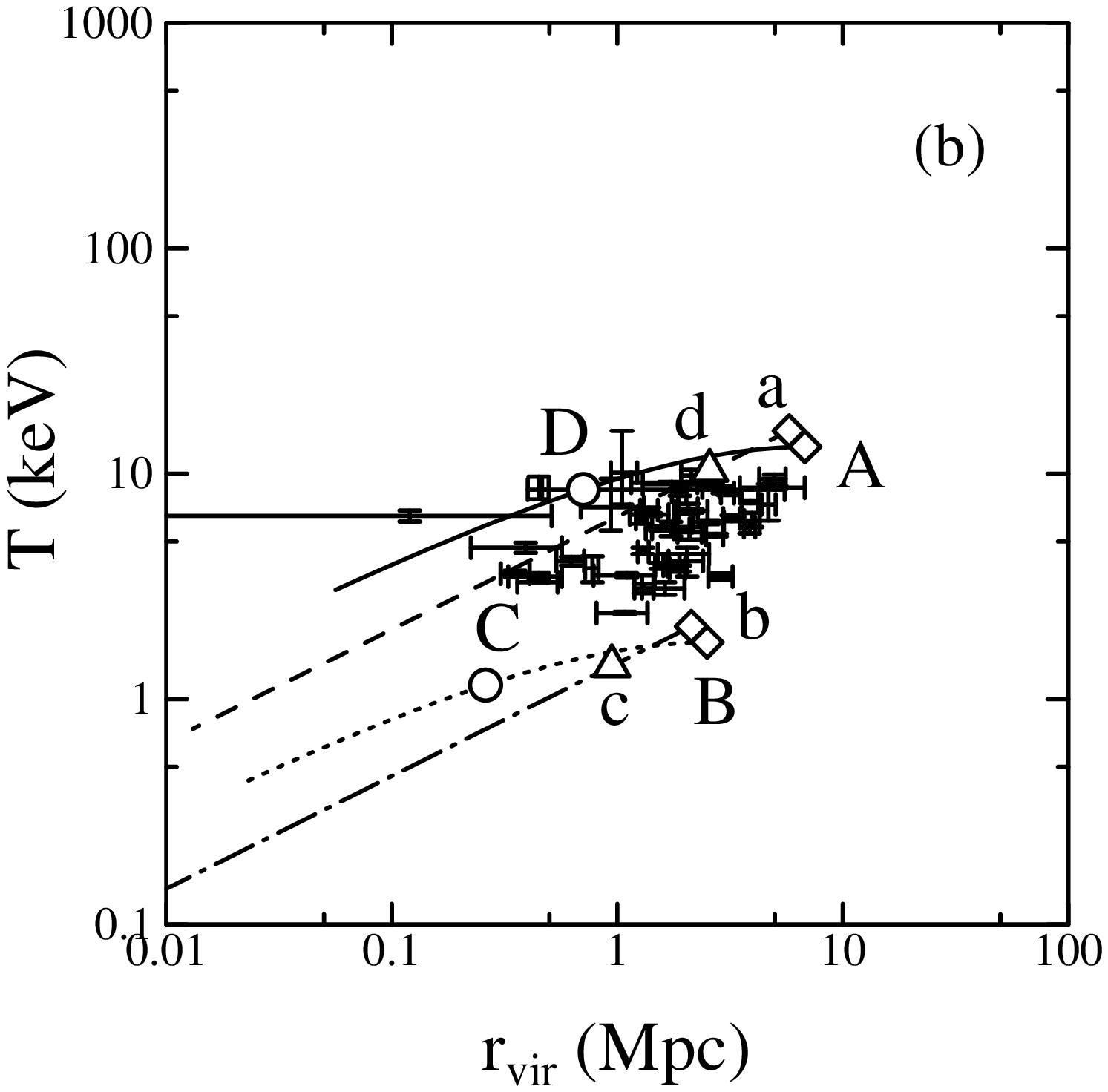, width=7cm}
\centering \epsfig{figure=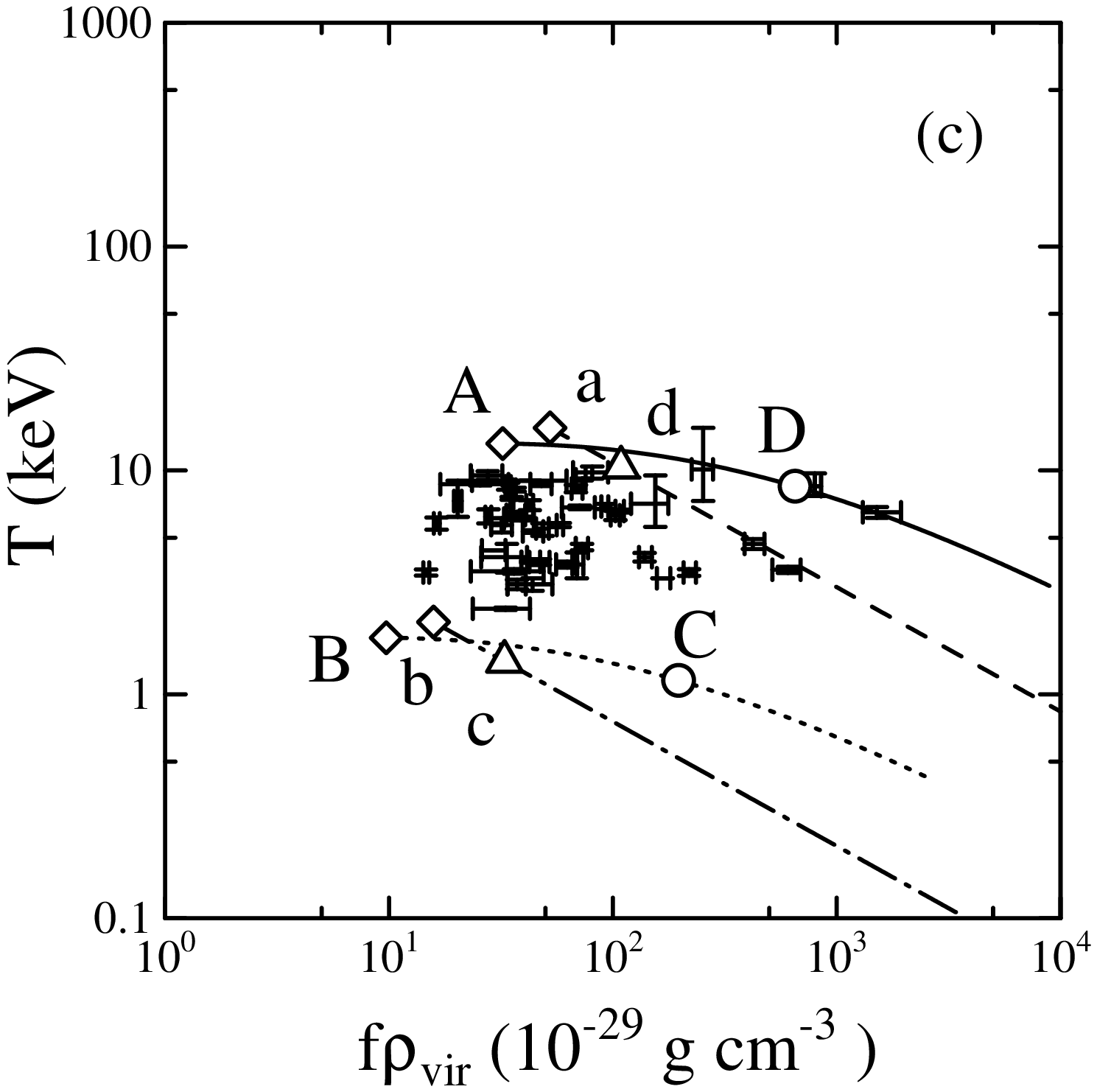, width=7cm} 
\caption{Theoretical predictions. (a) Radius--density relation (b)
radius--temperature relation (c) density--temperature relation. Solid
line: $\Omega_0=0.2$ and $M_{\rm vir}(t_0)=10^{16} \rm M_{\sun}$. Dotted
line: $\Omega_0=0.2$ and $M_{\rm vir}(t_0)=5\times 10^{14} \rm
M_{\sun}$. Dashed line: $\Omega_0=1.0$ and $M_{\rm vir}(t_0)=10^{16} \rm
M_{\sun}$. Dash-dotted line: $\Omega_0=1.0$ and $M_{\rm
vir}(t_0)=5\times 10^{14} \rm M_{\sun}$. The open diamonds (A, B, a, and
b), circles (C and D), and triangles (c and d) correspond to $z_{\rm
coll}=0$, $z_{\rm col}=1/\Omega_0-1$, and $z_{\rm col}=0.5$,
respectively. The observational data ($\rho_0$, $R$, and $T$) are
overlaid being shifted moderately in the directions of $\rho_0$ and $R$
($\rho_{\rm vir}=0.06\rho_0$ and $r_{\rm vir}=8R$).
} 
\end{figure}

\newpage

\begin{figure}
\centering \epsfig{figure=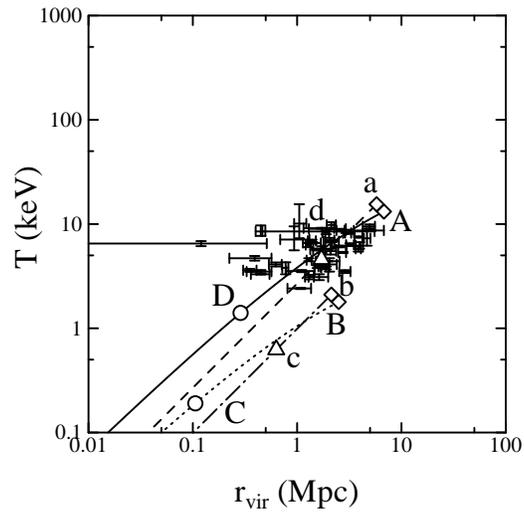, width=7cm} 
\caption{The same as Figure 1b, but for $n= -2$.} 
\end{figure}

\end{document}